\documentclass[pdflatex,sn-mathphys-num]{sn-jnl}

\usepackage{graphicx}%
\usepackage{multirow}%
\usepackage{amsmath,amssymb,amsfonts}%
\usepackage{amsthm}%
\usepackage{mathrsfs}%
\usepackage[title]{appendix}%
\usepackage{xcolor}%
\usepackage{textcomp}%
\usepackage{manyfoot}%
\usepackage{booktabs}%
\usepackage{algorithm}%
\usepackage{algorithmicx}%
\usepackage{algpseudocode}%
\usepackage{listings}%

\theoremstyle{thmstyleone}%
\newtheorem{theorem}{Theorem}
\theoremstyle{thmstyletwo}%

\theoremstyle{thmstylethree}%

\begin{document}

\title[Article Title]{Framework for Solving Quadratic Unconstrained Optimization Problem on a Gate Architecture}


\author*[1]{\fnm{Vladimir} \sur{Yatsulevich}}\email{vovas53@yandex.ru}

\author[1]{\fnm{Viktor} \sur{Lobachev}}\email{VLOB1969@yahoo.com}

\author[2]{\fnm{Anna} \sur{Ksel}}\email{ka053384@gmail.com}

\author[1,2]{\fnm{Yaroslav} \sur{Kholodov}}\email{ya.kholodov@innopolis.ru}

\affil*[1]{\orgname{Moscow Independent Research Institute of Artificial Intelligence (MIRAI)}, \orgaddress{\city{Moscow}, \postcode{117218}, \country{Russia}}}

\affil[2]{\orgdiv{Laboratory of Quantum Computing}, \orgname{Innopolis University}, \orgaddress{\city{Innopolis}, \postcode{420500}, \state{Tatarstan Republic}, \country{Russia}}}


\abstract{One of the promising areas in quantum computing today is quantum optimization. Many applied optimization problems, such as the courier distribution problem, protein docking, portfolio optimization, and others, can be effectively solved using quantum optimization algorithms. Quantum computers today represent the noisy intermediate-scale quantum era. This means that they are quite noisy, i.e., they can produce inaccurate results, and they contain a small number of qubits (on the order of tens or a few hundred). The main result of this work is a framework for solving optimization problems. To develop the framework, explicit formulas for constructing a quantum circuit based on a given Quadratic Unconstrained Binary Optimization matrix were obtained in three different basis sets. In addition, two methods for initializing the initial values of the variational parameters were proposed: trotterized quantum annealing variation and neural network approach. As part of the framework testing, it was demonstrated that the neural network approach to initializing variational parameters works better than the standard initialization. Additionally, as part of the work, an estimate was obtained of the number of program runs required to reach the global minimum. The proposed structure of framework is modular and allows for the integration of various methods for constructing quantum circuits, parameter initialization strategies, and classical optimization algorithms. This facilitates the systematic evaluation and comparison of variational quantum algorithms for solving binary optimization problems.}




\maketitle

\section{Introduction}

Discrete optimization is one of the most important areas of applied mathematics. A huge number of problems can be solved using discrete optimization methods. These include logistics problems. For example, the problem of distributing orders among couriers \cite{toth2002vehicle}. Another relevant area in the field of combinatorial optimization is the optimization of an equity portfolio. There are different formulations of this problem \cite{lu2024quantum}. Methods of discrete optimization have also found application in some bioinformatics tasks \cite{yanagisawa2024qubo}.

The main problem in discrete (or combinatorial) optimization lies in the inefficiency of existing methods. Many problems require a complete search of all possible solutions, which is exponentially complex. This means that many applied problems with large dimensions cannot be solved on classical computers.

Quantum algorithms make it possible to efficiently solve combinatorial problems. These include the QAOA (Quantum Approximate Optimization Algorithm) \cite{farhi2014quantumapproximateoptimizationalgorithm} and the QUBO (Quadratic Unconstrained Binary Optimization) model \cite{glover2011polynomial1, glover2011polynomial2}. Theoretically, quantum algorithms should provide incredible acceleration, provided that stable quantum computers are available. Quantum computers are currently at the stage of the Noise Intermediate Scale Quantum (NISQ) era \cite{preskill2018quantum}. This means that, firstly, quantum computers are small, i.e., they contain several dozen or hundreds of qubits. Secondly, the qubits in these computers are not ideal. This means that operations on them can only be performed with a certain degree of accuracy. And as the number of operators increases, the accuracy will decrease. It is also important to take into account the architecture of quantum processors.

Currently, there is a large number of methods and algorithms that could be used to solve specific optimization problems. Therefore, there is a need to develop a framework in which various methods could be tested. Thus, the goal of this work is to develop a set of software algorithms for solving combinatorial optimization problems using emulators or quantum processors that are currently available.
Another important task is to estimate the number of launch iterative procedure required to reach the global minimum. It is assumed that, at present, the expected dimensionality of the optimization problem will correspond to several dozen qubits.

The article will be organized as follows.
In Section 2. Preliminary Information, all the necessary notations will be presented. Section 3. Quantum Circuit Construction will be devoted to the algorithms for constructing a quantum circuit based on a given QUBO matrix. This section will also present an analysis of the constructed quantum circuits. Section 4. The selection of variational parameters will contain a description of the variation of trotterized quantum annealing and a neural network that predicts the initial values of the variational parameters.
In section 5. Framework, the framework, its constituent modules, as well as the description and results of the conducted experiments, will be presented. In section 6. Discussion, the further prospects for the development of each of the approaches presented in this work will be described.

\section{Preliminaries}\label{preliminaries}

\subsection{Quantum gates}\label{quantum_gates}

A qubit is a vector $|\psi\rangle$ in the space $\mathcal{H} = (\mathbb{C}^2)^{\otimes^n}$. Let us fix the standard computational basis $|\mathbf{x}\rangle = |x_1\rangle \otimes |x_2\rangle \otimes \dots \otimes |x_n\rangle$, where $ \forall i \in \{1, \dots, n\} \ x_i \in \{0, 1\}$. For further reasoning, the following set of unary operators will be used. First and foremost, this is the set of Pauli matrices

\begin{equation*}
    X = 
    \begin{pmatrix}
        0 & 1\\
        1 & 0
    \end{pmatrix}, \
    Y = 
    \begin{pmatrix}
        0 & -i\\
        i & 0
    \end{pmatrix},\ 
    Z = 
    \begin{pmatrix}
        1 & 0\\
        0 & -1
    \end{pmatrix},
\end{equation*}
Hadamard matrix
\begin{equation*}
    H = \frac{1}{\sqrt{2}}
    \begin{pmatrix}
        1 & 1\\
        1 & -1
    \end{pmatrix},
\end{equation*}
and the rotation operators $R_x, R_z$ around the $x, z$ axes on the Bloch sphere, respectively, by an angle $\theta$.
\begin{align*}
    R_x(\theta) &= 
    \begin{pmatrix}
        \cos\frac{\theta}{2} &  -i\sin\frac{\theta}{2}\\
        -i\sin\frac{\theta}{2} & \cos\frac{\theta}{2}
    \end{pmatrix}, 
    R_z(\theta) = 
    \begin{pmatrix}
        e^{-i\frac{\theta}{2}} & 0\\
        0 & e^{i\frac{\theta}{2}}
    \end{pmatrix}.
\end{align*}
A generalized rotation operator around an arbitrary axis will also be used.
\begin{equation*}
    U_3(\theta, \varphi, \lambda) = 
    \begin{pmatrix}
        \cos\frac{\theta}{2} & -e^{i\lambda}\sin\frac{\theta}{2}\\
        e^{i\varphi}\sin\frac{\theta}{2} & e^{i(\varphi + \lambda)}\cos\frac{\theta}{2}
    \end{pmatrix}.
\end{equation*}
Further in the text, the upper index in parentheses will indicate the number of the qubit to which this operator will be applied. For example,
\begin{equation*}
    X^{(j)}
    =
    \underbrace{I \otimes \dots \otimes I}_{j-1}
    \otimes X \otimes
    \underbrace{I \otimes \dots \otimes I}_{n-j}.
\end{equation*}

As binary operators, we will use controlled operators such as $C\text{NOT}$ and $CZ$. The operator will also be useful.

\begin{equation*}
    R_{zz}(\theta) = \exp\left(-i\frac{\theta}{2}Z \otimes Z\right).
\end{equation*}

\subsection{Discrete optimization}\label{discrete_optimization}

Many discrete optimization problems can be reduced to the QUBO model. Let $\mathbf{x} \in \{0, 1\}^n$ be a binary vector of dimension $n$, and let $Q$ be a symmetric or upper triangular real matrix of size $n \times n$. Let the functional in QUBO form be defined as follows
\begin{equation}\label{eq:qubo_obj_func}
    f(\mathbf{x}) = \mathbf{x}^TQ\mathbf{x}.
\end{equation}
Then the optimization problem in the QUBO form will take the form
\begin{equation*}
    \mathbf{x^*} = \underset{\mathbf{x} \in \{0, 1\}^n}{\text{argmin}}\  \mathbf{x}^T Q\mathbf{x}.
\end{equation*}

One of the main problems in the form of QUBO is the Max-Cut problem. Let a graph $G = (V, E)$ be given. It is necessary to partition the set of vertices into two disjoint sets $V = V_1 \sqcup V_2$ so that the number of edges of the form $(u, v) \in E$ $u \in V_1, v \in V_2$ is maximized.

In addition to problems in the form of QUBO, there are higher‑order problems — HUBO (High-ordered Unconstrained Binary Optimization).
Such a problem can either be reduced to QUBO, but with an increase in dimensionality, or solved directly. In the latter case, it is convenient to represent the functional as a tensor.

The Quantum Approximate Optimization Algorithm was first presented in the paper \cite{farhi2014quantumapproximateoptimizationalgorithm}. Let a binary functional be given
\begin{equation*}
    C: \{0, 1\}^n \rightarrow \mathbb{R}.
\end{equation*}
The Hamiltonian, in which the functional $C(\mathbf{x})$ will be encoded, has the form
\begin{equation*}
    \hat{H}_P = \sum_{\mathbf{x} \in \{0, 1\}^n}C(\mathbf{x})|\mathbf{x}\rangle\langle\mathbf{x}|.
\end{equation*}
The idea behind the algorithm is as follows. First, all qubits are in the state $|0\rangle$. Then, the Hadamard operator $H$ is applied to all qubits to put them into a superposition state.
\begin{equation*}
    |+\rangle^{\otimes^n} = \frac{1}{\sqrt{2^n}}\sum_{\mathbf{x} \in \{0, 1\}^n} |\mathbf{x}\rangle.
\end{equation*}
Further, instead of writing $|+\rangle^{\otimes^n}$, the shorter notation $|+\rangle$ will be used.
The operators are then defined
\begin{equation*}
    U_{\hat{H}_0}(\beta) = e^{-i\beta\hat{H}_0}, \ U_{\hat{H}_P}(\gamma) = e^{-i\gamma\hat{H}_P},
\end{equation*}
where $\beta \in [0, \pi]$, $\gamma \in [0, 2\pi]$, $\hat{H}_0 = \sum_{i = 1}^n X^{(i)}$.  Next, a natural number $p \in \mathbb{N}$ is fixed, and the vectors $\gamma = (\gamma_1, \dots, \gamma_p)$ and $\beta = (\beta_1, \dots, \beta_p)$ are defined. The final quantum circuit has the following form

\begin{equation}\label{eq:gamma_beta}
    |\gamma, \beta\rangle = U_{\hat{H}_0}(\beta_p)U_{\hat{H}_P}(\gamma_p)\dots U_{\hat{H}_0}(\beta_1)U_{\hat{H}_P}(\gamma_1)|+\rangle.
\end{equation}

It is claimed that for large $p$, there exist such $\gamma$ and $\beta$ that measuring the expectation value of the observable $\hat{H}_P$ for the state $|\gamma, \beta\rangle$ yields the minimum energy of the Hamiltonian $\hat{H}_P$. The QAOA algorithm is iterative. First, some initial value of the variational parameters $\gamma^{(0)}, \beta^{(0)}$ is selected. On a quantum computer, the state (\ref{eq:gamma_beta}) is calculated and a measurement is performed. Using some optimization algorithm, new values $\gamma^{(i)}, \beta^{(i)}$ are determined. This continues until the stopping conditions are met. This could be a proximity criterion or a maximum number of iterations.

It is important to note the connection between the QUBO model and the QAOA algorithm. Since the description of the QAOA algorithm uses an arbitrary binary functional $C(\mathbf{x})$, there exists a Hamiltonian $\hat{H}_P$ whose minimum energy corresponds to the solution of the QUBO problem. Let the QUBO functional be given in the form (\ref{eq:qubo_obj_func}); the expectation value of the quantum observable $\hat{H}_P$ in the state $|\mathbf{x}\rangle$ is as follows.
\begin{equation}\label{eq:quantum_objective}
    F(|\mathbf{x}\rangle) = \langle\mathbf{x} | \hat{H}_P |\mathbf{x}\rangle,
\end{equation}
and the value $Q_{jk}$ is an element of the QUBO matrix. Then, if
\begin{equation}\label{eq:ham_cost}
    \hat{H}_P = \sum_{j, k = 1}^{n}Q_{jk}\left(\frac{I - Z^{(j)}}{2}\right)\left(\frac{I - Z^{(k)}}{2}\right),
\end{equation}
then
\begin{equation*}
    \forall  \mathbf{x} \in \{0, 1\}^n \ f(\mathbf{x}) = F(|\mathbf{x}\rangle).
\end{equation*}
This result was proven in the paper \cite{hadfield2021representation}.


\section{Constructing a quantum circuit}\label{quantum_circuit}

The formula for constructing a quantum circuit for the Max-Cut problem is presented in the article \cite{blekos2024review}. In this work, a formula for composing a quantum circuit for an arbitrary QUBO matrix in various operator sets was derived. The corresponding result is presented in Theorem \ref{thm1}.

\begin{theorem}\label{thm1}
Let $Q$ be a QUBO matrix of dimension $n \times n$, and let the Hamiltonian be given in the form (\ref{eq:ham_cost}). Then the operator $U_{\hat{H}_P}(\gamma)$ is equivalent to a quantum circuit.
\begin{equation}\label{eq:qc_standart_form}
     \prod_{j,k=1}^{n} R_Z^{(j)}\left(-\frac{\gamma}{2}Q_{jk}\right)R_Z^{(k)}\left(-\frac{\gamma}{2}Q_{jk}\right)R^{(jk)}_{ZZ}\left(\frac{\gamma}{2}Q_{jk}\right).
\end{equation}
\end{theorem}
\begin{proof}
    To prove this theorem, it is necessary to explicitly substitute the Hamiltonian in the form (\ref{eq:ham_cost}) into the matrix exponential $e^{-i\gamma \hat{H}_P}$. After that, formally prove the commutativity of the operators in the powers. Break the exponential of the sum into a product of exponentials. And for each exponential, construct a unitary operator.
    
A detailed derivation of the explicit formula for constructing a quantum circuit can be found in Appendix \ref{appendix:quantum_circuit}. 
\end{proof}

Each factor within a large product can be conveniently visualized on a quantum circuit. Let $\alpha = -\gamma/2 \cdot Q_{jk}$.

\begin{figure}[h!]
\centering
\includegraphics{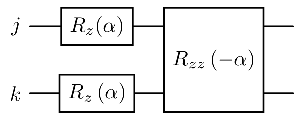}
\caption{One component of the operator $e^{-\gamma\hat{H}_P}$ of the QAOA quantum circuit in the $R_z, R_{zz}$ basis.}\label{img:qc_rz_rzz}
\end{figure}

Let’s estimate the number of operators. In formula (\ref{eq:qc_standart_form}), there are a total of $n^2$ factors, where each factor contains 3 operators. Then, the entire quantum circuit for the QAOA algorithm will require $\mathcal{O} ((3n^2 + n)p + 2n)$ operators, where $3n^2p$ is the number of gates for the operator $U_{\hat{H}_P}(\gamma)$, $np$ is the number of gates for the operator $U_{\hat{H}_0}(\beta)$, and the term $2n$ corresponds to $n$ Hadamard operators at the beginning of the quantum circuit and $n$ measurement operators at the end of the quantum circuit.

It is also useful to make an upper bound on the depth of the quantum circuit. The depth of a single operator $U_{\hat{H}_P}(\gamma)$ corresponds to two layers. The depth of the operator $U_{\hat{H}_0}(\beta)$ corresponds to one layer. Therefore, the upper bound will take the form $\mathcal{O}((2n^2 + 1)p + 2)$.

As a rule, quantum computers do not operate with the $R_{zz}$ operator, so it can be easily replaced with a combination of the $R_z$ and $C$NOT gates.
\begin{equation*}
    R_{zz}^{(jk)}(\alpha) = C\text{NOT}^{(jk)}R_z^{(k)}(\alpha)C\text{NOT}^{(jk)}.
\end{equation*} 
Without taking into account the phase factor, formula (\ref{eq:qc_standart_form}) can be rewritten as follows
\begin{equation}\label{eq:qc_cnot_form}
    \prod_{j,k=1}^{n} R_z^{(j)}\left(\alpha\right)R_z^{(k)}\left(\alpha\right)CX^{(jk)}R_z^{(k)}\left(-\alpha\right)CX^{(jk)},
\end{equation}
where $CX$ is another notation for the $C$NOT operator, used for brevity. For this form, $\mathcal{O} ((5n^2 + n)p + 2n)$ gates will be required. The depth of such a quantum circuit will be $\mathcal{O}((4n^2 + 1)p + 2)$.

\begin{figure}[h!]
\centering
\includegraphics{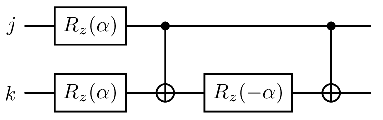}
\caption{One component of the operator $e^{-\gamma\hat{H}_P}$ of the QAOA quantum circuit in the $R_z, C$NOT basis.}\label{img:qc_rz_cnot}
\end{figure}

An important feature of the Wukong 180Q quantum computer \cite{originquantum} is that it works only with the $U_3$ and $CZ$ operators. Therefore, it becomes necessary to rewrite formulas (\ref{eq:qc_standart_form}) and (\ref{eq:qc_cnot_form}) using only these operators.
\begin{align}\label{eq:qc_u3_cz_form}
    \begin{split}
        \prod_{j,k=1}^{n}&U_3^{(j)}(0, \alpha, 0)U_3^{(k)}\left(\frac{\pi}{2}, \alpha, \pi \right)CZ^{(jk)}U_3^{(k)}\left(-\alpha, -\frac{\pi}{2}, \frac{\pi}{2}\right)CZ^{(jk)}U_3^{(k)}\left(\frac{\pi}{2}, 0, \pi\right).
    \end{split}
\end{align}

\begin{proof}
    You can find a detailed derivation of this formula in the Appendix \ref{appendix:quantum_circuit}. 
\end{proof}

The graphical representation of the quantum circuit is presented in Appendix \ref{appendix:quantum_circuit} in Fig. \ref{img:qc_u3_cz}. For this form, $\mathcal{O} ((6n^2 + n)p + 2n)$ gates will be required, or $\mathcal{O} ((11n^2 + n)p + 2n)$, if we assume that the quantum computer performs rotations around the $x, y, z$ axes as separate operations.

The depth of such a quantum circuit will be $\mathcal{O}((5n^2 + 1)p + 2)$. Again, if we consider all rotations to be sequential and independent, the depth increases to $\mathcal{O}((11n^2 + 1)p + 2)$.

\begin{figure}[h!]
\centering
\includegraphics[width=0.6\columnwidth]{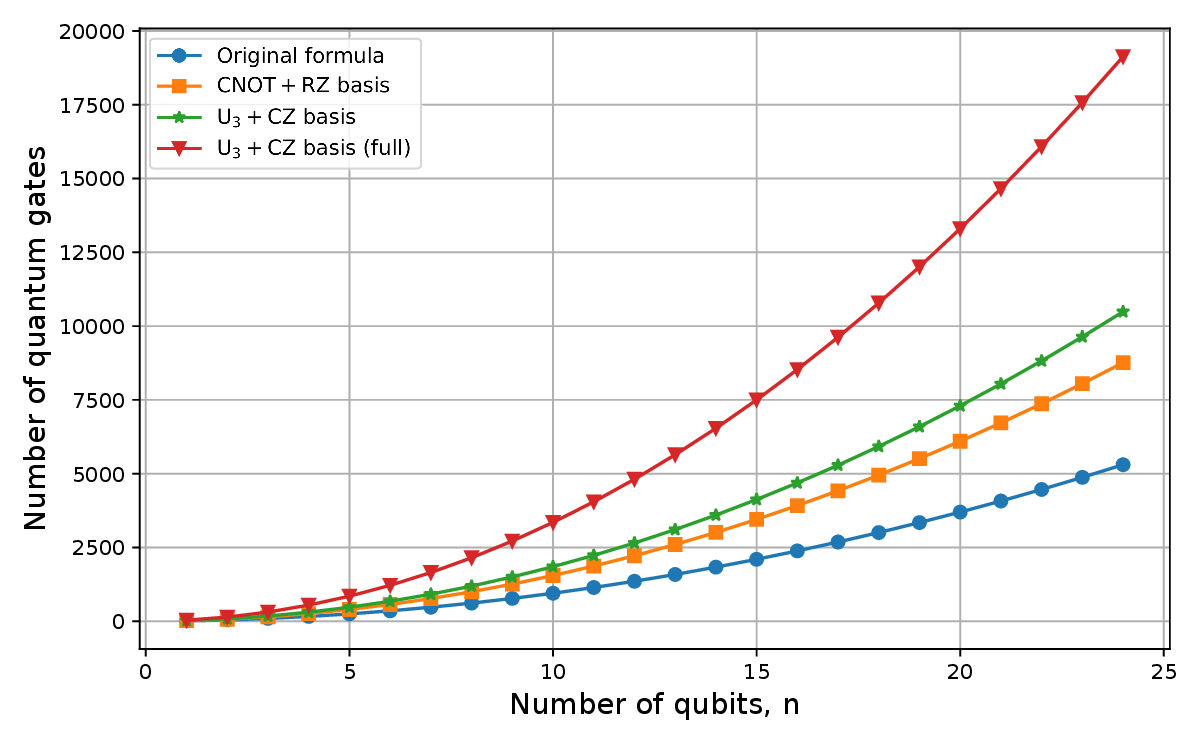}
\caption{Comparison of the number of quantum gates versus the dimensionality of the system ($p = 3$). The horizontal axis indicates the number of qubits. The vertical axis indicates the number of gates in the quantum circuit.}\label{fig:quantum_gates_comparing}
\end{figure}

Figure \ref{fig:quantum_gates_comparing} shows a graph of the dependence of the number of quantum gates on the dimensionality of the system.

\begin{figure}[h!]
\centering
\includegraphics[width=0.6\columnwidth]{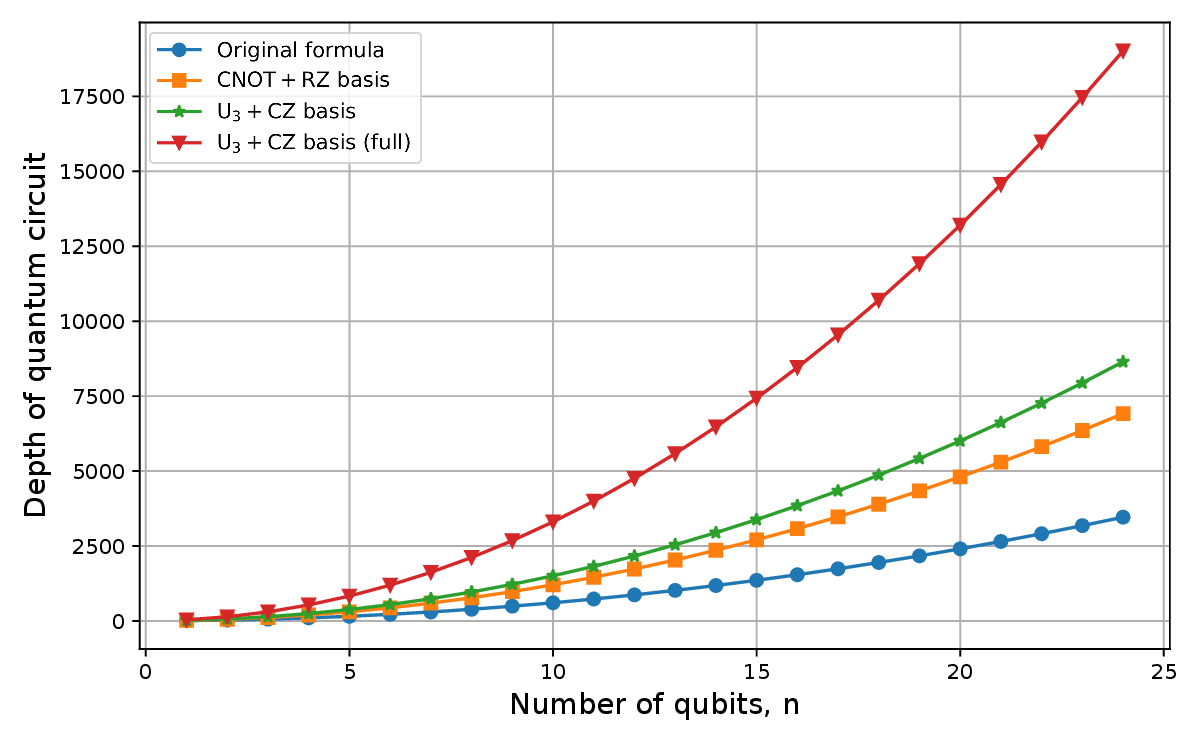}
\caption{Comparison of the depth of a quantum circuit versus the dimensionality of the system ($p = 3$). The number of qubits is indicated on the horizontal axis. The depth of the quantum circuit is indicated on the vertical axis.}\label{fig:quantum_circuit_depth_comparing}
\end{figure}

Figure \ref{fig:quantum_circuit_depth_comparing} shows the relationship between the depth of a quantum circuit and the dimensionality of the system. The depth of a quantum circuit is a more significant indicator than the total number of operators, as it allows for an estimate of the maximum number of qubits that can be used on available hardware with a given level of quality. For example, the Wukong 180Q quantum computer allows running quantum circuits with a depth of no more than 500 \cite{originquantum}. This means that at present, it is possible to run a task with a dimensionality of no more than 5.

\section{Estimation of variation parameters}

The main challenge of the QAOA algorithm lies in determining the optimal values of the variational parameters. Here, two main problems can be identified. First, one could simply iterate through all possible values of the variational parameters on a certain lattice. However, even with a small lattice, the number of calls to the quantum computer will be high, while computers in the NISQ era do not allow for such a large number of runs. In addition, the fixed resolution of the lattice does not guarantee the determination of the global solution due to the high level of oscillations in the energy landscape. The second global problem is the choice of an optimization algorithm. As mentioned earlier, the energy landscape is generally highly oscillatory, which means the number of local minima will be high. Standard gradient methods do not work, so there is a need to develop alternative algorithms. The best results from optimization algorithms are achieved with the correct choice of initial values for the variational parameters this is the so‑called warm‑start.

\subsection{Trotterized quantum annealing}\label{tqa}

One of the methods for initializing the initial values of variational parameters is based on the physical interpretation of the QAOA algorithm. This variational algorithm is a discrete version of quantum annealing. In quantum annealing, the energy operator is constructed as follows.
\begin{equation*}
    \hat{H}(t) = \left(1 - \frac{t}{T}\right)\hat{H}_0 + \frac{t}{T}\hat{H}_P,
\end{equation*}
where $T$ is a sufficiently large number that depends on the original QUBO matrix \cite{ohzeki2011quantum}. A unitary operator is constructed based on this Hamiltonian, and the Trotter–Suzuki formula is applied \cite{farhi2014quantumapproximateoptimizationalgorithm, trotter1959product}. As a result, we obtain exactly the same set of operators as in the standard formulation of the QAOA algorithm, only with specific values of the variational parameters. Based on this logic, it is possible to determine the initial values of the variational parameters, which may be close to optimal. This approach is called Trotterized quantum annealing TQA \cite{sack2021quantum}.

One of the results of this work is a variation of the TQA algorithm. In the work \cite{aharonov2008adiabatic}, an estimate of the adiabatic evolution time was given. For an arbitrary $\varepsilon > 0$, we fix some $\delta > 0$. Then the adiabatic evolution time will be lower-bounded as follows
\begin{equation*}
    T \geqslant \Omega\left( \frac{\|\hat{H}_P - \hat{H}_0\|^{1 + \delta}}{\varepsilon^\delta \min\limits_{s\in [0, 1]} \{\Delta^{2 + \delta}\hat{H}(s)\} } \right),
\end{equation*}
where $\|H\| \equiv \max_w\frac{\|Hw\|}{\|w\|}$, and $\Delta \hat{H}_P$ is the spectral gap of the Hamiltonian $\hat{H}_P$, and the resulting quantum state will be $\varepsilon$-close to the ground state of the Hamiltonian $\hat{H}_P$. Based on this inequality, an estimate of the initial values of the variational parameters of the QAOA algorithm was obtained. The full statement of Theorem \ref{thm2} and its proof is presented in Appendix \ref{appendix:physical_intuition}.

\subsection{Machine learning approach}\label{ml}

Another relevant method for initializing the initial values of variational parameters is the use of machine learning methods. As a machine learning model, one can use both standard models, such as linear or trigonometric regression, and more complex ones, such as neural networks; moreover, the training can be either supervised or unsupervised. In this work, the choice fell on a multilayer perceptron.

In the paper \cite{jain2022graph}, the authors trained a graph neural network for an approximate solution to the Max-Cut problem. The approximate solution was encoded in a Hamiltonian, which was embedded in a QAOA chain. In the paper \cite{amosy2022iterative}, the authors trained a separate neural network for each dimension of the graph, which directly predicts the values of the variational parameters. In another study \cite{meng2025conditional}, the authors trained a diffusion neural network. Thus, research in this area is actively underway.

The dataset consists of a set of QUBO matrices constructed based on the Max-Cut problem with 5–8 vertices. For each graph, an exact solution was found by exhaustive search. Let the mapping be
\begin{equation*}
    \text{NN}: Q, \mathbf{w} \mapsto (\beta^*, \gamma^*)
\end{equation*}
it is a neural network that will predict the values of the variational parameters based on the QUBO matrix. In the formula, $\mathbf{w}$ represents the trainable parameters of the neural network. Further, let
\begin{equation*}
    \text{QC}: Q, (\beta, \gamma) \mapsto E.
\end{equation*}
This is a function that will run a quantum computing simulator and output the average energy. Then, as a loss function, we can consider a function of the form
\begin{equation*}
    \text{loss}(Q, \mathbf{w}) = \frac{\text{QC}(Q, \text{NN}(Q, \mathbf{w})) - E_{\min}(Q)}{E_{\max}(Q) - E_{\min}(Q)}.
\end{equation*}

\begin{figure}[h!]
\centering
\includegraphics[width=0.5\columnwidth]{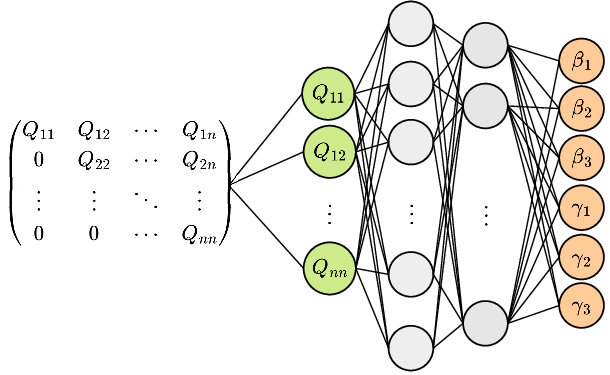}
\caption{The architecture of the neural network predicting the values of the variational parameters. The input layer consists of 36 neurons. The upper‑triangular matrix at the input is transformed into a linear layer. There are two hidden layers: the first contains 128 neurons, the second contains 64 neurons. The output layer contains 6 neurons, which corresponds to the 6 values of the variational parameters when $p = 3$.}\label{fig:perceptron}
\end{figure}

As a result, a model was trained, which was integrated into the framework’s initialization module. The results of the neural network’s operation are presented in the next section. The architecture of the neural network is shown in Fig. \ref{fig:perceptron}.

It is important to note that the training of a neural network begins with the initialization of the initial weight values. These values were determined randomly. The predictions of the given neural network depend on the initial initialization. For this reason, several neural networks were trained for different initial weight initialization methods. The model with the best quality metric performance was added to the framework.

It is assumed that the neural network approach will make it possible to find the optimal values of the variational parameters without an iterative process. On the other hand, neural networks will be able to localize the region in which further search will be carried out. This will also reduce the number of iterations.


\section{Framework}\label{framework}

The QAOA algorithm was introduced more than 10 years ago. And during this time, several software systems have been developed that can be used to solve various discrete optimization problems. As a rule, all solutions are Python libraries. Meanwhile, our developed framework is a platform that allows you to solve optimization problems without writing code. Such platforms include \verb|OpenQuantumComputing| \cite{fuchs2024qaoa}, \verb|PennyLane|
\cite{bergholm2018pennylane} and \verb|QiskitOptimization| \cite{qiskit_optimization}. In addition to the fact that there is no need to write code, an important advantage is the sequential assembly of quantum code without using standard quantum computing libraries, while \verb|Qiskit| and \verb|Cirq|
are used as simulators that run the already assembled quantum code. The architecture of our software allows adding the implementation of arbitrary variational quantum algorithms.

\subsection{Framework description}\label{framework_description}

As part of the research, software was developed that is aimed at solving discrete optimization problems. This framework is presented as a platform in a no-code format. This means that users of this application will not need to write code to implement the quantum optimization algorithm.

The fundamental operation scheme of the framework is as follows. The program receives a QUBO matrix for a certain problem as input. Then, based on the QUBO matrix and certain values of the variational parameters, a quantum code is constructed in the \verb|QASM| language \cite{cross2017open}. This code is sent to a quantum computing emulator or to a physical quantum processor. As a result of the emulation or direct quantum computing, the average energy value is determined. The found energy value and the variational parameters are passed to a certain optimizer.

\begin{figure}[h!]
\centering
\includegraphics[width=0.6\columnwidth]{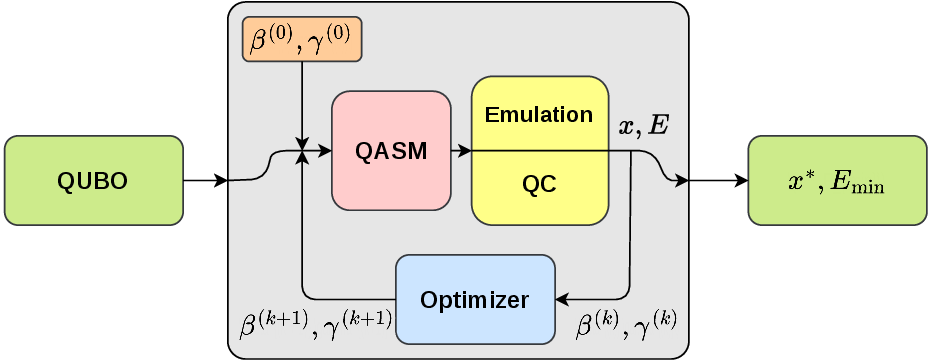}
\caption{The diagram of the QAOA algorithm. The input and output data are highlighted in green. The orange block represents the initial value of the variational parameters; the red block is the block for constructing quantum code in QASM (Quantum Assembler); the yellow block corresponds to quantum computing (emulator or quantum computation (QC)); the blue block is the optimizer.}\label{fig:qaoa}
\end{figure}

This diagram describes only the general picture of how the framework works and does not cover all of its functionality. The following will provide a detailed description of all the functional blocks of the framework.

\textbf{Input data}. The program will receive a certain tensor as input, which will encode the discrete optimization problem. First and foremost, this is a QUBO matrix. In addition to standard quadratic optimization problems, the framework provides the ability to solve higher‑order problems.

\textbf{Variational algorithm}. It is important to note that the software complex is designed so that the user has the opportunity to select the variational algorithm they need. Accordingly, depending on the selected algorithm, the schemes for constructing the variational parameters will differ.

\begin{figure}[h!]
\centering
\includegraphics[width=0.6\columnwidth]{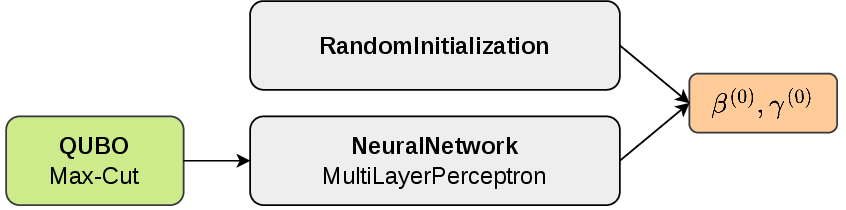}
\caption{The diagram of the initialization block. Currently, the framework supports two types of initializers: a standard random one and a neural network. The neural network is capable of accepting QUBO matrices with a size ranging from 5 to 8 as input.}\label{fig:initialization}
\end{figure}

\textbf{Initialization of variational parameters}. One of the most important areas of research in the field of variational algorithms is determining the initial values of the variational parameters. The standard approach involves random initialization. This approach is not effective, since even in the case of a single-layer QAOA algorithm, the number of local minima can be quite large. This paper proposes a method for determining the initial values of the variational parameters based on the physical process of quantum annealing. This result is presented in Theorem \ref{thm2}. This approach has not yet been implemented and requires additional research. The third approach is to initialize the values of the variational parameters using a neural network. This block incorporates a neural network that has been previously trained on a large dataset and is capable of predicting these values. The block for initializing the values of the variational parameters is shown in Fig. \ref{fig:qaoa} as an orange rectangle. A more detailed representation of the initialization block is given in Fig. \ref{fig:initialization}.

\begin{figure}[h!]
\centering
\includegraphics[width=0.65\columnwidth]{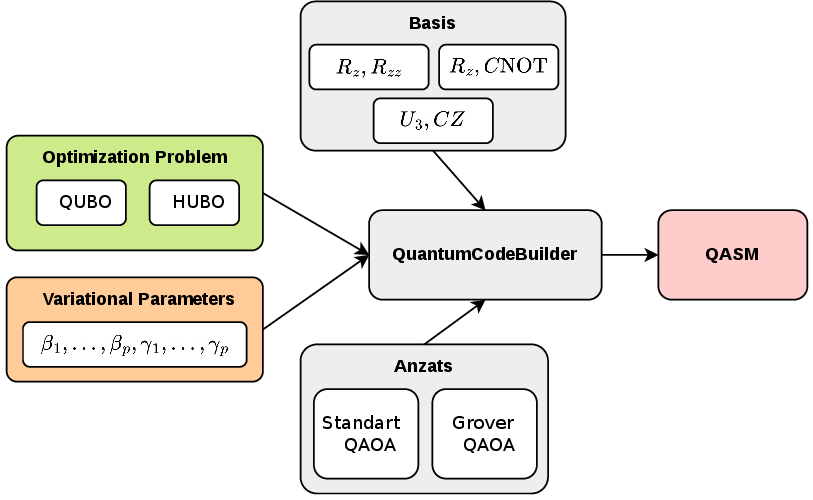}
\caption{The QuantumCodeBuilder block takes as input an optimization problem in the form of QUBO or HUBO, the values of the variational parameters, the version of the quantum algorithm, and the operator basis. As a result of processing, this block outputs quantum code written in the Quantum Assembler (QASM) programming language.}\label{fig:quantum_code_builder}
\end{figure}

\textbf{Quantum code construction}. The most important part of the framework is the module for constructing quantum code based on a given QUBO matrix and the specified values of the variational parameters. Constructing quantum code is a mapping of the form
\begin{equation*}
    \verb|QuantumCodeBuilder|: Q, (\beta, \gamma) \mapsto \verb|quantum_code.qasm|.
\end{equation*}
It is at this stage that the formulas (\ref{eq:qc_standart_form}), (\ref{eq:qc_cnot_form}), and (\ref{eq:qc_u3_cz_form}) are used. Three different operator bases are embedded in the software module to enable execution on various quantum processors. The software architecture allows for expanding the quantum code construction schemes by adding new sets of operators. This block corresponds to the red rectangle in Fig. \ref{fig:qaoa}. A more detailed representation of the quantum code assembly block is shown in Fig. \ref{fig:quantum_code_builder}.

It is important to note that at the current stage, the quantum code is reassembled at each iteration of the optimization process, which is not an effective approach. In the future, it is planned to develop a system for precompiling quantum code, followed by substitution of the values of variational parameters. With this approach, the quantum code template will be created once for a given optimization problem. And at each iteration of the optimization process, the final code assembly will be performed by substituting the variational parameters.

\textbf{Quantum Computing}. This software module is the core that is directly responsible for calculations at the quantum level. In the NISQ era, quantum computers are not widely used, so this software module will include both physical quantum processors and quantum computing simulators. Currently, the core includes the quantum computing simulators \verb|Cirq| \cite{cirq} and \verb|Qiskit| \cite{javadi2024quantum}. On Fig. \ref{fig:qaoa} The quantum computing module corresponds to the yellow rectangle.

\textbf{Optimizer}. The next important module is the optimizer itself. This module is highlighted in blue in Fig. \ref{fig:qaoa}. It is a classic algorithm that, based on the results of the quantum computing module, produces new values for the variational parameters. There are various optimizers: gradient-based and non-gradient-based. Currently, various non-gradient methods from the \verb|nevergrad| library \cite{nevergrad} are used, as well as the COBYLA optimizer \cite{Powell1994directsearch}.

\textbf{Compression of quantum code}. One of the main problems with directly constructing a quantum code is the large number of elementary operations. As shown in Fig. \ref{fig:quantum_circuit_depth_comparing}, the depth of a quantum circuit grows very rapidly even for relatively small $n \leqslant 25$. Moreover, while for emulators this affects only the execution time, for quantum computers it introduces obstacles in the form of qubit coherence times and errors at each operator. After all, even with extremely high quality $\approx0.9999$ at a depth of $10^4$, the result of the algorithm turns into randomness. Therefore, the issue of quantum code compression is extremely acute. Currently, two approaches are expected to be used: the \verb|Qiskit| \cite{javadi2024quantum} compressor and the \verb|PyZX| \cite{kissinger2019pyzx} approach.

\textbf{Use cases}. Implementing only the standard version of the algorithm is not significant. Therefore, it is important to implement various experimental variations of the variational algorithm within the framework. For example, there is an iterative approach called INTERP \cite{sack2021quantum}, in which the number of layers in the QAOA algorithm is gradually increased. Another interesting work is recursive QAOA. Researchers in the paper \cite{sack2023recursive} presented an analytical result stating that, given a known local minimum on $p$ layers, it is possible to guarantee obtaining a new local minimum on $p + 1$ layers, the energy of which will be guaranteed to be lower.

\subsection{Experiments}\label{experiment}

To demonstrate the functionality of the developed framework, the following experiment was conducted. The following parameters were recorded. Quantum computing emulator~--- \verb|Qiskit|; number of measurements: 1000; operator basis~--- $CZ, U_3$; optimizer~--- \verb|OnePlusOne|; number of optimization steps: 100; initialization of variational parameters: random.

The metric of quality chosen was the achievement of the global minimum among the 5 most likely measurement results. The idea is to evaluate not the average energy, but the number of times the global minimum is achieved over 100 runs of the iterative process. This metric will allow us to estimate the number of runs in the actual use of the framework.

\begin{figure}[h!]
\centering
\includegraphics[width=0.55\columnwidth]{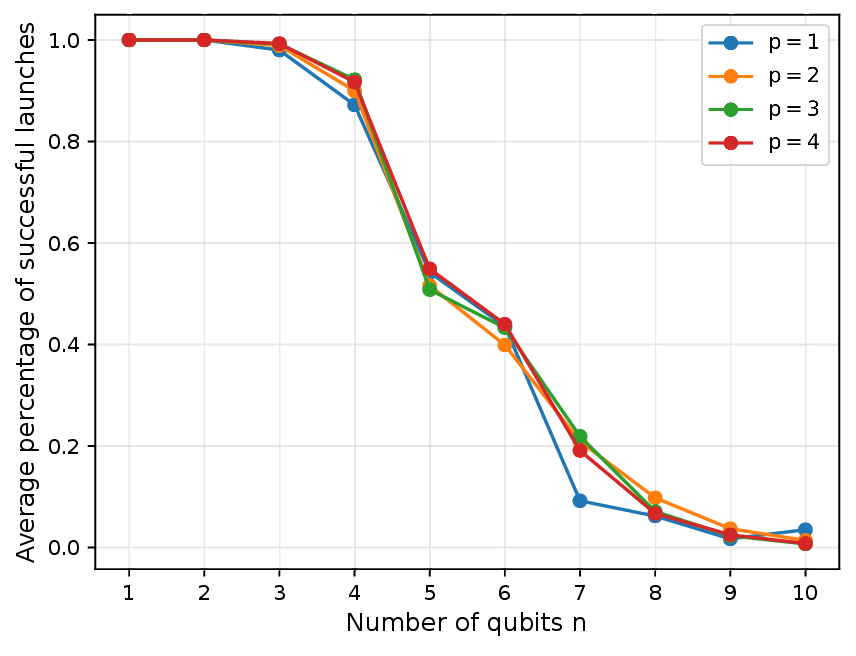}
\caption{The result of the framework’s operation using the standard QAOA algorithm with random initialization of the variational parameters. The horizontal axis shows the number of qubits. The vertical axis shows the average percentage of successful runs. A successful run is defined as achieving the global minimum in at least one of the five most likely solutions.}\label{fig:exp}
\end{figure}

The measurements were taken for matrices with dimensions ranging from $1$ to $10$. For each size, 10 random QUBO matrices were generated, and the exact solution was determined using brute‑force search. Then, the QAOA algorithm was run on each matrix with $p = 1, 2, 3, 4$ layers. As a result, the average probability of reaching the global minimum was determined for the given parameters $(n, p)$. The experimental results are presented in Fig. \ref{fig:exp}.

From the results of the standard algorithm, it is clear that the unmodified version can only be used to solve discrete optimization problems in the form of QUBO with a small dimensionality. Or, in other words, a larger number of experiment runs is required to solve optimization problems with large dimensions.

Let us estimate the number of experiment runs. Let $p_{\text{succ}}$ be the probability of conducting a successful experiment, i.e., the probability of reaching the global minimum, and let $N$~ be the number of runs required to reach the global minimum. Then the probability of success in at least one of the repetitions will be equal to
\begin{equation*}
    P = 1 - (1 - p_{\text{succ}})^N.
\end{equation*}
Let $\varepsilon$ be the permissible error, that is, the probability of not obtaining the global minimum in $N$ runs. Then the inequality must hold
\begin{equation*}
    1 - (1 - p_{\text{succ}})^N \geqslant 1 - \varepsilon.
\end{equation*}
By taking the logarithm of the expression, it is easy to obtain an estimate of the number of runs
\begin{equation*}
    N_{\min} = \left\lceil \frac{\ln \varepsilon}{\ln (1 - p_{\text{succ}})} \right\rceil.
\end{equation*}

From the graph in Fig. \ref{fig:exp}, it is clear that the inflection point in the average probability of success occurs at approximately $n \approx 5$. A good approximation of the data in the graphs is achieved using the function
\begin{equation*}
    p_{\text{succ}}(n) = \frac{1}{1 + e^{n - n_0}},
\end{equation*}
where $n_0$ can be taken as the fracture size, i.e., $n_0 = 5$. The pattern of the probability dependence on the number of layers in the QAOA algorithm at small dimensions is rather weak, so it will not be considered at the current stage. For the sake of correctness of the reasoning, let’s fix $p = 3$. Then the estimate of the number of runs will take the form.
\begin{equation*}
    N_{\min} = \left\lceil \frac{\ln \varepsilon}{\ln \left(\frac{e^{n - n_0}}{1 + e^{n - n_0}}\right)} \right\rceil.
\end{equation*}

Fig. \ref{fig:exp} shows an experiment conducted with random initialization of the initial values of the variation parameters. Experiments on the TQA variation were not carried out, as the obtained expressions require a separate in‑depth analysis. A separate experiment was conducted for the neural network. The dimensions $n = 5\dots 8$, $p = 3$ were fixed. For each dimension, 10 random matrices for the Max‑Cut problem were selected. Measurements were taken in the same way as in the first experiment.

\begin{figure}[h!]
\centering
\includegraphics[width=0.45\columnwidth]{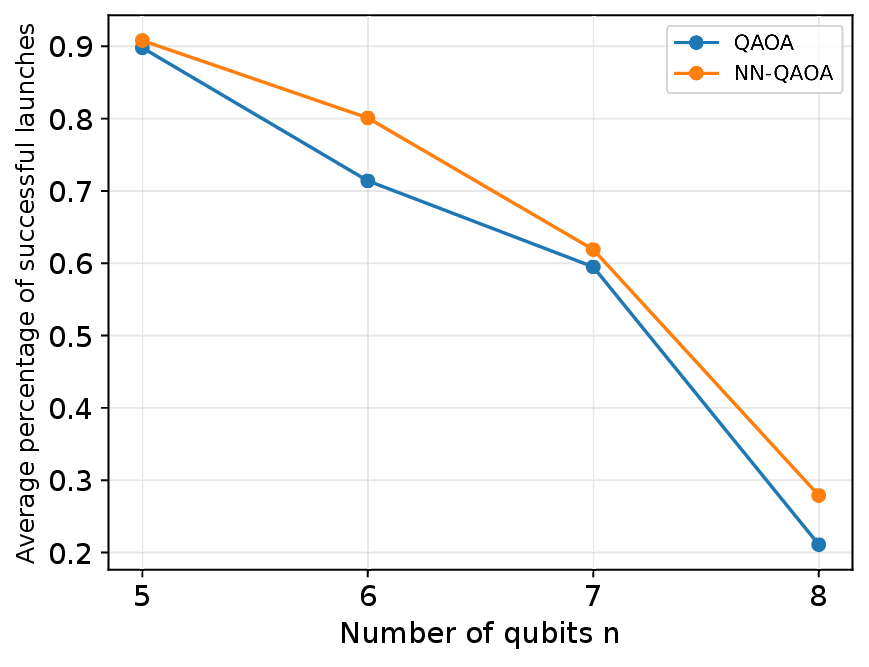}
\caption{The result of the framework’s operation using the QAOA algorithm with initial values of the variational parameters initialized using a neural network. The number of qubits is indicated on the horizontal axis. The average percentage of successful runs is indicated on the vertical axis. A successful run is defined as achieving a global minimum in at least one of the five most likely solutions.}\label{fig:exp_nn}
\end{figure}

The results of the experiment with the neural network are presented in Fig. \ref{fig:exp_nn}. Based on the graph, it is clear that the neural network provides a slight advantage compared to the standard QAOA algorithm. It is important to note that the model built is a demonstration model, as it is applicable only to dimensions 5-8. Nevertheless, this result proves the possibility of applying machine learning approaches for warm-start. It is worth noting that the performance of the variational algorithm is on a comparable level with existing works. For example, the work \cite{lotshaw2021empirical} shows similar results for the Max-Cut problem. As part of this study, it was important to assess the probability of achieving the global minimum for an arbitrary QUBO matrix, and this was done.


\section{Discussion}\label{discussion}

The main result of the work is a framework for solving optimization problems. Its main advantage is that there is no need to write code. The second main advantage of the framework is its modularity. Each module is a set of various methods and algorithms. Every year, papers are published in which new methods are developed and can be integrated as part of the framework. The framework has this capability. That is, if a group of researchers develops a new method, they can easily integrate it into the framework without changing its overall architecture. Among the framework’s drawbacks is the compilation of quantum code at each iteration of the optimization process, which is not efficient. In the future, it is planned to optimize this software module.

In addition to the framework, the work presented explicit formulas for quantum circuits and warm-start methods. Explicit formulas were presented only for the standard QAOA and only for problems in QUBO form. In the future, it is planned to implement quantum circuits for problems of arbitrary order. The presented warm-start methods also require additional research. In the case of a TQA variation, it is necessary to add either an assessment of the spectrum of the Hamiltonian in which the problem is encoded to the initialization module, or to perform optimizations using a different set of parameters, and then conduct experiments. The advantage of the current version of the TQA variation lies in the use of an estimate of the minimum spectral gap.

Initializing variational parameters using machine learning models is an extremely relevant area of research at present. The results obtained prove the viability of ML algorithms. Even a relatively simple model is capable of predicting the values of variational parameters. This indicates that a more flexible architecture is needed to produce a more accurate result. As an example, we can consider graph neural networks. Their distinctive feature lies in the data structure. A problem in QUBO form can be naturally reduced to graphs. Therefore, graph neural networks are the most suitable candidate for the role of an optimal and efficient architecture.


\section{Conclusion}\label{conclusion}

As part of the work, a software package was developed to solve optimization problems in the QUBO form. To implement one of the modules, explicit formulas were derived for constructing a quantum circuit based on a given matrix in various operator sets. Two methods for initializing the values of the variational parameters were also proposed: a variation of Trotterized quantum annealing and a neural network. Two experiments were conducted: with the standard QAOA and with the neural network. Both experiments demonstrated the functionality of both the software package and the presented methods. The number of runs needed to reach the global minimum was also estimated, and this is a quite significant outcome for the practical use of the framework. In the future, the framework is planned to be expanded. Various optimizers, quantum code compressors, and much more will be added to it.


\backmatter

\bmhead{Supplementary information}\label{addition}

During the work, software for conducting experiments was developed. The source code can be found in the GitHub repository at the link \cite{github}.

\begin{oldappendices}


\section{Constructing a quantum circuit}\label{appendix:quantum_circuit}

\textbf{Basis $R_z, R_{zz}$}

Let’s start with the proof of the main statement presented in Theorem \ref{thm1}.

\begin{proof}
    In the original paper \cite{farhi2014quantumapproximateoptimizationalgorithm}, the unitary operator in which the optimization problem is encoded is of the form $U_{\hat{H}_P}(\gamma) = e^{-i\gamma\hat{H}_P}$, where $\gamma \in [0, 2\pi]$ is a variational parameter, and the Hamiltonian encodes a QUBO matrix:
    \begin{equation}\label{eq:ham_appendix}
        \hat{H}_P = \sum_{j, k = 1}^{n}Q_{jk}\left(\frac{I - Z^{(j)}}{2}\right)\left(\frac{I - Z^{(k)}}{2}\right).
    \end{equation}
    Accordingly, first it is necessary to substitute the Hamiltonian (\ref{eq:ham_appendix}) into the operator $U_{\hat{H}_P}(\gamma)$
    \begin{equation}
        \exp \left(-\frac{i\gamma}{4} \sum_{j,k=1}^{n} Q_{jk} \left(I-Z^{(j)}\right)\left(I -Z^{(k)}\right)\right).
    \end{equation}
    Furthermore, it can be observed that operators of the form $(I-Z^{(j_1)})(I-Z^{(k_1)})$ and $(I-Z^{(j_2)})(I-Z^{(k_2)})$ commute. This means that the exponential of the sum can be split into a product of exponentials
    \begin{align}
        \prod_{j,k=1}^{n} \exp \left(-\frac{i\gamma}{4}Q_{jk} \left(I-Z^{(j)}\right)\left(I-Z^{(k)}\right) \right).
    \end{align}
    After expanding the brackets inside the product, it is also necessary to note that the operators $I - Z^{(j)}$ and $I - Z^{(k)}$ commute. Let $\alpha_{jk} = (-i\gamma/4)
    \cdot Q_{jk}$. Then
    \begin{align}
        \begin{split}
            \prod_{j,k=1}^{n} \exp\left(\alpha_{jk}I\right)\exp\left(-\alpha_{jk}Z^{(j)}\right)\exp\left(-\alpha_{jk}Z^{(k)}\right)\exp\left(\alpha_{jk}Z^{(j)}Z^{(k)}\right).
        \end{split}
    \end{align}
    Now let’s analyze each factor. The factor of the form $\exp\left(\alpha_{jk}I\right)$ is a phase factor, which means it does not affect the final distribution. Thus, the first factor can be removed.
    The second operator is related to the unitary rotation gate $R_z$ as follows
    \begin{equation}
        \exp \left(-\frac{i\gamma}{4}Q_{jk} Z_j\right) =  R_z^{(j)}\left(\frac{\gamma}{2}Q_{jk} \right).
    \end{equation}
    The last factor can be expressed in terms of the operator $R_{zz}$
    \begin{equation}
        \exp\left(\frac{i\gamma}{4}Q_{jk}Z_jZ_k\right) = R_{zz}^{(jk)}\left(-\frac{\gamma}{2}Q_{jk}\right)
    \end{equation}
    By combining the presented expressions, we obtain formula (\ref{eq:qc_standart_form}).
\end{proof}

\textbf{Basis $U_3, CZ$.}

\begin{proof}
    To prove the correctness of formula (\ref{eq:qc_u3_cz_form}), it is sufficient to specify a few operator relations. First, we need to express the $C$NOT operator in terms of $CZ$.
\begin{equation}
    C\text{NOT}^{(jk)} = H^{(k)} \cdot CZ^{(jk)} \cdot H^{(k)}.
\end{equation}
Next, we need to express the Hadamard operator in terms of the rotation operator
\begin{equation}
    H = U_3\left(\pi/2, 0, \pi\right).
\end{equation}
First, we will make a direct replacement of the operators in the formula (\ref{eq:qc_cnot_form})
\begin{equation}\label{eq:pred_u3_cz}
    R_z^{(j)}(\alpha)R_z^{(k)}(\alpha)H^{(k)}CZ^{(jk)}H^{(k)}R_z^{(k)}(-\alpha)H^{(k)}CZ^{(jk)}H^{(k)}.
\end{equation}
The rotation operator $R_x$ is related to the rotation operator $R_z$ by the following relationship
\begin{equation}
    R_x(\alpha) = HR_z(\alpha)H = U_3(\alpha, -\pi/2, \pi/2).
\end{equation}
The rotation operator $R_z(\alpha)$ is related to the operator $U_3$ by the following relationship
\begin{equation}
    R_z(\alpha) = e^{i\frac{\alpha}{2}} U_3(0, \alpha, 0).
\end{equation}
There is one more operator relationship left
\begin{equation}
    R_z(\alpha)H = U_3(\pi/2, \alpha, \pi).
\end{equation}
All that remains is to substitute the operators into equation (\ref{eq:pred_u3_cz}).
\end{proof}

\begin{figure}[h!]
\centering
\includegraphics[width=0.7\columnwidth]{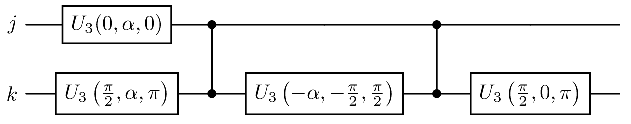}
\caption{One component of the operator $e^{-\gamma\hat{H}_P}$ of the QAOA quantum circuit in the $U_3, CZ$ basis.}\label{img:qc_u3_cz}
\end{figure}


\section{Variational parameters tuning using physical foundation of quantum annealing}\label{appendix:physical_intuition}

\begin{theorem}\label{thm2}
    Let the QUBO matrix $Q$ be given. Then, to run the QAOA algorithm, the following estimates can be taken as the initial values of the hyperparameters.
    \begin{align*}
        p &=  \left\lceil\frac{|\| \hat{H}_P\| - n|}{\varepsilon - \|r(\delta s)\|}\right\rceil,\\
        \gamma_k &= \left[ C \cdot\frac{|n - \|\hat{H}_P\|^\delta (\varepsilon - \|r(\delta s)\|) |}{\eta \cdot \varepsilon^{\delta} \min \{ 1, \Delta^{2 + \delta}\hat{H}_P\}}\cdot\left(1 - k\delta s\right)\right]_{2\pi},\\
    \beta_k &= \left[ C \cdot \frac{k (\varepsilon - \|r(\delta s)\|)^2}{\eta \cdot \varepsilon^\delta \min \{ 1, \Delta^{2 + \delta}\hat{H}_P\} \cdot |\|\hat{H}_P\| - n|^{1 - \delta}}\right]_{2\pi},
    \end{align*}
    where $\|r(\delta t)\| \xrightarrow{} 0 $ if $\delta t \xrightarrow{} 0$, $[x]_{2\pi} = x - 2\pi \lfloor x / 2\pi \rfloor$ and $\delta s = (\varepsilon - \|r(\delta s)\|) / (|\| \hat{H}_P\| - n|)$.
\end{theorem}
 
\begin{proof}
    To begin with, the Hamiltonian of adiabatic evolution is expressed in a general form as follows
\begin{equation}
    \hat{H}(s) = A(s)\hat{H}_0 + B(s)\hat{H}_P,
\end{equation}
where $A(0) = 1, A(1) = 0$ and $B(0) = 0, B(1) = 1$. We will base our work on the paper \cite{aharonov2008adiabatic}, specifically Theorem \ref{thm2}.1. Let us fix $A(s) = 1 - s$, and $B(s) = s$. We will determine $\Delta(\hat{H}(s))$ — the difference between the two minimum energy levels. For an arbitrary $\varepsilon > 0$, we fix some $\delta > 0$. Then the time of adiabatic evolution will be bounded from below as follows
\begin{equation}
    T \geqslant \Omega\left( \frac{\|\hat{H}_P - \hat{H}_0\|^{1 + \delta}}{\varepsilon^\delta \min\limits_{s\in [0, 1]} \{\Delta^{2 + \delta}\hat{H}(s)\} } \right),
\end{equation}
where $\|\hat{H}_P\| \equiv \max_w\frac{\|\hat{H}_Pw\|}{\|w\|}$. The value of time $T$ directly affects the values of $A(s)$ and $B(s)$, as well as the partition $\delta t$.

Let’s start with the estimation of time $T$. In the numerator, we will use the triangle inequality $\|\hat{H}_P - \hat{H}_0\| \geqslant |\|\hat{H}_P\| - \|\hat{H}_0\||$. For $\hat{H}_0$, the spectral norm is known. For a system of $n$ qubits, $\|\hat{H}_0\| = n$. The complexity arises in estimating the norm $\|H_P\|$. The problem is that the exact value of the norm $\|\hat{H}_P\|$ corresponds to the solution of the QUBO problem. At a minimum, $\|\hat{H}_P\| \geqslant \mathbf{x}^TQ\mathbf{x}$. For example, consider the following estimate. Let $\mathbf{x} = (1, 1, \dots, 1)$. Then $\mathbf{x}^TQ \mathbf{x} = \sum_{j, k} Q_{jk}$. Therefore, $\|H_P\| \geqslant \sum_{j,k}Q_{jk}$. This estimate is rather weak and requires refinement. Other estimates can also be used.

Now let’s try to estimate the value of $\Delta(\hat{H}(s))$. The strategy will be as follows. We will estimate the gap for the operator $\hat{H}_0$ and for the operator $\hat{H}_P$. Then
\begin{equation}
    \min\limits_{s\in [0, 1]} \{\Delta^{2 + \delta}\hat{H}(s)\} \geqslant \eta \cdot \min\{\Delta^{2 + \delta}\hat{H}_0, \Delta^{2 + \delta}\hat{H}_P\},
\end{equation}
where $\eta > 0$ is a sufficiently small coefficient. 

\begin{figure}[h!]
\centering
\includegraphics[width=0.5\columnwidth]{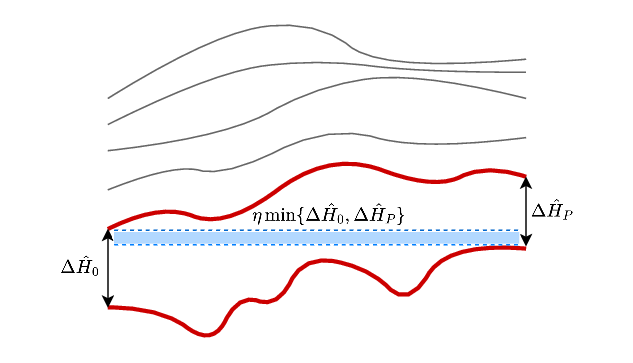}
\caption{Approximation of the minimum spectral gap using a line whose width, for any time $t$, will be less than the spectral gap $\hat{H}(s)$.}\label{fig:spectral_gap_approximation}
\end{figure}

The minimum spectral gap is approximated by the height of the widest line that can be inscribed between the first two energy levels of the operator $\hat{H}(s)$.
For the operator $\hat{H}_0$, the eigenvalues are of the form $\lambda = \sum\limits_{i = 1}^n \lambda_i$, where $\lambda_i = \pm 1$. This means that the distance between two adjacent energy levels is equal to two.

Taking into account all of the above, the estimate of the time required for adiabatic evolution will take the following form.
\begin{equation}\label{eq:B4}
    T \geqslant C\cdot\frac{|n - \|\hat{H}_P\||^{1+\delta}}{\eta \cdot\varepsilon^\delta \min\{1, \Delta^{2 + \delta}\hat{H}_P\}},
\end{equation}
where $C > 0$ is some constant obtained as a result of expanding the $\Omega$-notation.

Now let’s move on to the discretization of the algorithm. Let’s divide the time $T$ into $p$ intervals $\delta t$. To preserve adiabaticity in the discrete version, it is important to maintain the hierarchy of energy levels. This can be done as follows. Consider the Hamiltonian at some moment in time $t_k = k \delta t$, and $s_k = t_k / T$, where $1 \leqslant k \leqslant p$.
\begin{equation}
    \hat{H}(s_k) = A(s_k)\hat{H}_0 + B(s_k)\hat{H}_P.
\end{equation}
Now, let’s add a small change in the interval $\delta s = \delta t / T$ to the point $s_k$.
\begin{equation}
    \hat{H}(s_k + \delta s) = A(s_k + \delta s)\hat{H}_0 + B(s_k + \delta s)\hat{H}_P.
\end{equation}

The task is to estimate $\delta t$ and determine $p = T/\delta t$. Let’s start by defining the Fréchet derivative of the operator $H$.
\begin{align}
    \hat{H}(s_k + \delta s) - \hat{H}(s_k) = L(s_k)\delta s + r(\delta s),
\end{align}
where $L(s) = A'(s)\hat{H}_0 + B'(s)\hat{H}_P$ is the Frechet derivative of the operator $\hat{H}(s)$, and $r(\delta s) = \alpha(\delta s)\hat{H}_0 + \beta(\delta s)\hat{H}_P$ is the operator residual term, where $\alpha(\delta s)$ and $\beta(\delta s)$ are infinitesimal functions corresponding to the operators $\hat{H}_0$ and $\hat{H}_P$, respectively. Let us evaluate the residual term $r(\delta s)$
\begin{equation}
    \|r(\delta s)\| \leqslant |\alpha(\delta s)|\|\hat{H}_0\| + |\beta(\delta s)|\|\hat{H}_P\|.
\end{equation}
Let’s divide the resulting expression by $\delta s$
\begin{equation}
    \frac{\|r(\delta s)\|}{\delta s} = \frac{|\alpha(\delta s)|}{\delta s}\|\hat{H}_0\| + \frac{|\beta(\delta s)|}{\delta s}\|\hat{H}_P\|.
\end{equation}
It is important to note that the order of the functions $\alpha$ and $\beta$ is no less than two. This means that as $\delta s \rightarrow 0$, the values $|\alpha(\delta s)|/\delta s$ and $|\beta(\delta s)|/\delta s$ will also tend to zero. And this means
\begin{equation}
    \frac{\|r(\delta s)\|}{\delta s} \rightarrow 0.
\end{equation}

Let us proceed to the evaluation of the operator‑valued function $\hat{H}(s)$. It is necessary that, for some $\varepsilon > 0$, the difference in the spectrum be less than $\varepsilon$, i.e.
\begin{equation}
    |\|\hat{H}(s_k)\| - \|\hat{H}(s_k + \delta s)\|| < \varepsilon.
\end{equation}
Let’s start the estimation.
\begin{align}
   \|L(t_k)\delta t + r(\delta t)\| \leqslant \delta t \|L(t_k)\| + \|r(\delta t)\|.
\end{align}
If we set the value obtained by the expression to be less than $\varepsilon$, then the chain of inequalities will be satisfied.
\begin{equation}
    |\|\hat{H}(s_k)\| - \|\hat{H}(s_k + \delta s)\|| < \varepsilon.
\end{equation}
Let’s express $\delta s$ from the previous chain.
\begin{equation}
    \delta s \|L(s_k)\| < \varepsilon - \|r(\delta t)\| \Leftrightarrow \delta s < \frac{\varepsilon - \|r(\delta s)\|}{\|L(s_k)\|}
\end{equation}
Consider that $A(s) = 1 - s$ and $B(s) =s$, then $A'(s)= -1$ and $B'(s) = 1$. That
\begin{equation}\label{eq:B15}
    \delta s < \frac{\varepsilon - \|r(\delta s)\|}{\|\hat{H}_P - \hat{H}_0\|} \leqslant \frac{\varepsilon - \|r(\delta s)\|}{|\|\hat{H}_P\| -n|}.
\end{equation}

In the original paper \cite{farhi2014quantumapproximateoptimizationalgorithm}, the author introduces the QAOA algorithm as a discrete version of quantum annealing. In this case, we can set the hyperparameter values to $\gamma_k = \delta t (1 - t_k/T)$, $\beta_k =\delta t\cdot t_k/T$, or moving on to the variables $s_k$, we get
\begin{equation}\label{eq:B16}
    \gamma_k = T\delta s (1 - k \delta s), \quad \beta_k = Tk (\delta s)^2. 
\end{equation}

Now, substituting the estimate for $\delta s$ and (\ref{eq:B15}) and for $T$ (\ref{eq:B4}) into the formula (\ref{eq:B16}), we get
\begin{align}
    \begin{split}
        \gamma_k &=\Bigg[C\cdot\frac{|n - \|\hat{H}_P\|^\delta (\varepsilon - \|r(\delta s)\|) |}{\eta \cdot \varepsilon^{\delta} \min \{ 1, \Delta^{2 + \delta}\hat{H}_P\}}\cdot\left(1 - k\cdot\frac{\varepsilon - \|r(\delta s)\|}{|\|\hat{H}_P\| - n|}\right)\Bigg]_{2\pi},\\
        \beta_k &= \left[C\cdot\frac{k (\varepsilon - \|r(\delta s)\|)^2}{\eta \cdot \varepsilon^\delta \min \{ 1, \Delta^{2 + \delta}\hat{H}_P\} \cdot |\|\hat{H}_P\| - n|^{1 - \delta}}\right]_{2\pi},
    \end{split}
\end{align}
where $[x]_{2\pi} = x - 2\pi \lfloor x / 2\pi \rfloor$. You can also make an estimate for the number of iterations of $p$
\begin{equation}
    p = \left\lceil\frac{1}{\delta s}\right\rceil = \left\lceil\frac{|\| \hat{H}_P\| - n|}{\varepsilon - \|r(\delta s)\|}\right\rceil.
\end{equation}
The explanation described above confirms the following Theorem.
\end{proof}

It is important to note that the results obtained cannot be considered exact solutions, since a large number of approximations were made in the reasoning. This approach can be useful as a parametrization of the $\gamma_k$ and $\beta_k$ parameters, since the number of optimized parameters can be reduced in the optimization algorithm. These parameters include the estimation of the spectral norm of $\|\hat{H}_P\|$, estimation of the spectral gap of the Hamiltonian in which the problem $\Delta\hat{H}_P$, approximation accuracy $\varepsilon, \delta$, spectral norm of the infinitesimal operator $\|r(\delta s)\|$, scalability coefficient of adiabatic evolution time $C$ and narrowing coefficient of the linear approximation of the spectral gap $\eta$. A total of 7 parameters are obtained. This approach will have a theoretical advantage for $p > 3$. An estimate of the spectral norm can be obtained based, for example, on the inverse relaxation of \cite{poljak1995convex}.

\end{oldappendices}


\bibliography{sn-bibliography}

@misc{farhi2014quantumapproximateoptimizationalgorithm,
  author        = {Farhi, Edward and Goldstone, Jeffrey and Gutmann, Sam},
  title         = {A Quantum Approximate Optimization Algorithm},
  year          = {2014},
  eprint        = {1411.4028},
  archivePrefix = {arXiv},
  primaryClass  = {quant-ph},
  doi           = {10.48550/arXiv.1411.4028},
}

@misc{cirq,
  author       = {{Google Quantum AI}},
  title        = {{Cirq}},
  howpublished = {Software repository},
  url          = {https://github.com/quantumlib/Cirq},
  note         = {Accessed 27 March 2026}
}

@article{aharonov2008adiabatic,
  author  = {Aharonov, Dorit and Van Dam, Wim and Kempe, Julia and Landau, Zeph and Lloyd, Seth and Regev, Oded},
  title   = {Adiabatic Quantum Computation Is Equivalent to Standard Quantum Computation},
  journal = {SIAM Rev.},
  volume  = {50},
  number  = {4},
  pages   = {755--787},
  year    = {2008},
  doi     = {10.1137/080734479},
}

@misc{originquantum,
  author       = {{Origin Quantum}},
  title        = {{Origin Pilot}: China's First Quantum Computer Operating System},
  howpublished = {Website},
  url          = {https://originqc.com.cn/en/},
  note         = {Accessed 26 April 2026}
}

@book{toth2002vehicle,
author = {Toth, Paolo and Vigo, Daniele},editor = {Paolo Toth and Daniele Vigo},
title = {The Vehicle Routing Problem},
publisher = {Society for Industrial and Applied Mathematics},
year = {2002},
doi = {10.1137/1.9780898718515},
address = {Philadelphia, PA 19104-2688},
}

@misc{cross2017open,
  author        = {Cross, Andrew W. and Bishop, Lev S. and Smolin, John A. and Gambetta, Jay M.},
  title         = {Open Quantum Assembly Language},
  year          = {2017},
  eprint        = {1707.03429},
  archivePrefix = {arXiv},
  doi           = {10.48550/arXiv.1707.03429},
}

@misc{nevergrad,
  author       = {Rapin, J. and Teytaud, O.},
  title        = {{Nevergrad}: A Gradient-Free Optimization Platform},
  year         = {2018},
  howpublished = {GitHub repository},
  url          = {https://github.com/facebookresearch/nevergrad}
}

@misc{lu2024quantum,
  author        = {Lu, Ying-Chang and Fu, Chao-Ming and Yu, Lien-Po and Chang, Yen-Jui and Chang, Ching-Ray},
  title         = {Quantum-Inspired Portfolio Optimization in the {QUBO} Framework},
  year          = {2024},
  eprint        = {2410.05932},
  archivePrefix = {arXiv},
  doi           = {10.48550/arXiv.2410.05932},
}

@article{yanagisawa2024qubo,
  author  = {Yanagisawa, Keisuke and Fujie, Takuya and Takabatake, Kazuki and Akiyama, Yutaka},
  title   = {{QUBO} Problem Formulation of Fragment-Based Protein--Ligand Flexible Docking},
  journal = {Entropy},
  volume  = {26},
  number  = {5},
  pages   = {397},
  year    = {2024},
  doi = {10.3390/e26050397}
}

@article{glover2011polynomial1,
  author  = {Glover, Fred and Hao, Jin-Kao and Kochenberger, Gary},
  title   = {{Polynomial Unconstrained Binary Optimisation --- Part 1}},
  journal = {Int. J. Metaheuristics},
  volume  = {1},
  number  = {3},
  pages   = {232--256},
  year    = {2011},
  doi = {10.1504/IJMHEUR.2011.041196}
}

@article{glover2011polynomial2,
  author  = {Glover, Fred and Hao, Jin-Kao and Kochenberger, Gary},
  title   = {{Polynomial Unconstrained Binary Optimisation --- Part 2}},
  journal = {Int. J. Metaheuristics},
  volume  = {1},
  number  = {4},
  pages   = {317--354},
  year    = {2011},
  doi = {10.1504/IJMHEUR.2011.044356}
}

@article{preskill2018quantum,
  author  = {Preskill, John},
  title   = {{Quantum Computing in the NISQ Era and Beyond}},
  journal = {Quantum},
  volume  = {2},
  pages   = {79},
  year    = {2018},
  doi = {10.22331/q-2018-08-06-79}
}

@article{ohzeki2011quantum,
  author  = {Ohzeki, Masayuki and Nishimori, Hidetoshi},
  title   = {Quantum Annealing: An Introduction and New Developments},
  journal = {J. Comput. Theor. Nanosci.},
  volume  = {8},
  number  = {6},
  pages   = {963--971},
  year    = {2011},
  doi     = {10.1166/jctn.2011.1776},
}

@article{poljak1995convex,
  author  = {Poljak, Svatopluk and Wolkowicz, Henry},
  title   = {Convex Relaxations of $(0,1)$-Quadratic Programming},
  journal = {Math. Oper. Res.},
  volume  = {20},
  number  = {3},
  pages   = {550--561},
  year    = {1995},
  doi     = {10.1287/moor.20.3.550},
}

@article{blekos2024review,
  author  = {Blekos, Kostas and Brand, Dean and Ceschini, Andrea and Chou, Chiao-Hui and Li, Rui-Hao and Pandya, Komal and Summer, Alessandro},
  title   = {A Review on Quantum Approximate Optimization Algorithm and Its Variants},
  journal = {Phys. Rep.},
  volume  = {1068},
  pages   = {1--66},
  year    = {2024},
  doi = {10.1016/j.physrep.2024.03.002}
}

@misc{github,
  author       = {{YatsulevichVV}},
  title        = {{QuantumOptimization}},
  howpublished = {GitHub repository},
  url          = {https://github.com/YatsulevichVV/QuantumOptimization},
  note         = {Accessed 15 September 2026}
}

@inbook{Powell1994directsearch,
  author = {Powell, M. J. D.},
  title  = {A Direct Search Optimization Method That Models the Objective and Constraint Functions by Linear Interpolation},
  year   = {1994},
  doi    = {10.1007/978-94-015-8330-5_4},
}

@misc{javadi2024quantum,
  author        = {Javadi-Abhari, Ali and Treinish, Matthew and Krsulich, Kevin and Wood, Christopher J. and Lishman, Jake and Gacon, Julien and Martiel, Simon and Nation, Paul D. and Bishop, Lev S. and Cross, Andrew W. and others},
  title         = {Quantum Computing with {Qiskit}},
  year          = {2024},
  eprint        = {2405.08810},
  archivePrefix = {arXiv},
  doi           = {10.48550/arXiv.2405.08810},
}

@article{sack2023recursive,
  author  = {Sack, Stefan H. and Medina, Raimel A. and Kueng, Richard and Serbyn, Maksym},
  title   = {Recursive Greedy Initialization of the Quantum Approximate Optimization Algorithm with Guaranteed Improvement},
  journal = {Phys. Rev. A},
  volume  = {107},
  number  = {6},
  pages   = {062404},
  year    = {2023},
  doi     = {10.1103/PhysRevA.107.062404},
}

@article{sack2021quantum,
  author  = {Sack, Stefan H. and Serbyn, Maksym},
  title   = {Quantum Annealing Initialization of the Quantum Approximate Optimization Algorithm},
  journal = {Quantum},
  volume  = {5},
  pages   = {491},
  year    = {2021},
  doi     = {10.22331/q-2021-07-01-491},
}

@misc{kissinger2019pyzx,
  author        = {Kissinger, Aleks and Van De Wetering, John},
  title         = {{PyZX}: Large Scale Automated Diagrammatic Reasoning},
  year          = {2019},
  eprint        = {1904.04735},
  archivePrefix = {arXiv},
  doi           = {10.48550/arXiv.1904.04735},
}

@article{trotter1959product,
  author  = {Trotter, Hale F.},
  title   = {On the Product of Semi-Groups of Operators},
  journal = {Proc. Am. Math. Soc.},
  volume  = {10},
  number  = {4},
  pages   = {545--551},
  year    = {1959},
  doi     = {10.1090/S0002-9939-1959-0108732-6},
}

@article{jain2022graph,
  author  = {Jain, Nishant and Coyle, Brian and Kashefi, Elham and Kumar, Niraj},
  title   = {Graph Neural Network Initialisation of Quantum Approximate Optimisation},
  journal = {Quantum},
  volume  = {6},
  pages   = {861},
  year    = {2022},
  doi     = {10.22331/q-2022-11-17-861},
}

@article{hadfield2021representation,
  author  = {Hadfield, Stuart},
  title   = {{On the Representation of Boolean and Real Functions as Hamiltonians for Quantum Computing}},
  journal = {ACM Trans. Quantum Comput.},
  volume  = {2},
  number  = {4},
  pages   = {1--21},
  year    = {2021},
  doi = {10.1145/3478519}
}

@article{amosy2022iterative,
  author  = {Amosy, Ohad and Danzig, Tamuz and Lev, Ohad and Porat, Ely and Chechik, Gal and Makmal, Adi},
  title   = {Iteration-Free Quantum Approximate Optimization Algorithm Using Neural Networks},
  journal = {Quantum Mach. Intell.},
  volume  = {6},
  number  = {2},
  pages   = {38},
  year    = {2024},
  doi     = {10.1007/s42484-024-00159-y},
}

@article{meng2025conditional,
  author  = {Meng, Fanxu and Zhou, Xiangzhen and Zhu, Pengcheng and Luo, Yu},
  title   = {Conditional Diffusion-Based Parameter Generation for Quantum Approximate Optimization Algorithm},
  journal = {EPJ Quantum Technol.},
  volume  = {12},
  number  = {1},
  pages   = {100},
  year    = {2025},
  doi     = {10.1140/epjqt/s40507-025-00397-4},
}

@software{fuchs2024qaoa,
  author  = {Fuchs, Franz Georg},
  title   = {{QAOA}: A Modular Python Library for the Quantum Approximate Optimization Algorithm},
  year    = {2024},
  url     = {https://github.com/OpenQuantumComputing/QAOA},
    note         = {Accessed 23 October 2026}
}

@misc{bergholm2018pennylane,
  author        = {Bergholm, Ville and Izaac, Josh and Schuld, Maria and Gogolin, Christian and Blank, Carsten and McKiernan, Keri and Killoran, Nathan},
  title         = {{PennyLane}: Automatic Differentiation of Hybrid Quantum-Classical Computations},
  year          = {2018},
  eprint        = {1811.04968},
  archivePrefix = {arXiv},
  doi           = {10.48550/arXiv.1811.04968},
}

@misc{qiskit_optimization,
  author       = {{Qiskit Optimization Development Team}},
  title        = {{Qiskit Optimization}},
  howpublished = {Software repository},
  url          = {https://github.com/qiskit-community/qiskit-optimization},
  note         = {Accessed 22 October 2026}
}

@article{lotshaw2021empirical,
  author  = {Lotshaw, Phillip C. and Humble, Travis S. and Herrman, Rebekah and Ostrowski, James and Siopsis, George},
  title   = {Empirical Performance Bounds for Quantum Approximate Optimization},
  journal = {Quantum Inf. Process.},
  volume  = {20},
  number  = {12},
  pages   = {403},
  year    = {2021},
  doi     = {10.1007/s11128-021-03298-0},
}

\end{document}